\documentclass[aps,groupedaddress,amssymb,amsmath,amsbsy,amsfonts,twocolumn,pra]{revtex4-1}

\usepackage{amssymb,amsmath,amsthm,color,graphicx,times,graphicx}
\usepackage{hyperref}
\usepackage{subfigure}
\usepackage{times}
\usepackage{bbold}
\usepackage{color}

\theoremstyle{plain}

\theoremstyle{definition}

\begin{document}

\title{Controlling Fano resonance and slow/fast light in a magnomechanical system with an optical parametric amplifier}
	
	\author{M'bark Amghar}
	\affiliation{LPTHE, Department of Physics, Faculty of Sciences, Ibnou Zohr University, Agadir, Morocco}
	\author{Noura Chabar}
	\affiliation{LPTHE, Department of Physics, Faculty of Sciences, Ibnou Zohr University, Agadir, Morocco}
	\author{Mohamed Amazioug}
	\affiliation{LPTHE, Department of Physics, Faculty of Sciences, Ibnou Zohr University, Agadir, Morocco}

\begin{abstract}

We study the slow-fast light effect and multi-transparency induced by magnomechanical systems. The system incorporates two magnons, which are collective magnetic excitations, placed alongside a degenerate optical parametric amplifier (OPA) within a cavity. The interaction between phonons, magnons, and light inside the cavity leads to two phenomena: magnomechanically induced transparency (MMIT) and magnon induced transparency (MIT). We show how an OPA alters the absorption and dispersion characteristics of the light spectrum. The observation of the Fano resonance through magnon-mechanical coupling with the degenerate OPA is discussed. Through tuning vibration interactions and the OPA, we achieve the improvement of slow light. We hope our findings could pave the way for advancements in quantum information processing.	

\end{abstract}
	\date{\today}

	\maketitle
	
\section{Introduction}
	
	The light-matter interaction has emerged as a significant and active research topic, with potential applications in various fields such as electromagnetically induced transparency (EIT) \cite{1,2,3}, photon and magnon blockade \cite{4,5,6,7}, quantum entanglement \cite{asjad,9,SUllah}, macroscopic quantum superposition states \cite{mabdi}, squeezing \cite{Jieli} and optomechanically induced transparency (OMIT) \cite{11,13,14,15,amghar}. Over the past decade, it has fostered advancements in robust quantum memory and the processing of quantum information. EIT is a quantum interference effect observed in three-level atoms \cite{16}. It occurs when the absorption of atoms can be suppressed to zero by an auxiliary laser field, which is primarily due to interference effects or the resonance of the dark state in an excited state. Additionally, phenomena analogous to EIT, resulting from the destructive interference between the weak probe field and the anti-Stokes scattering field, are called optomechanically induced transparency (OMIT) \cite{17}. This phenomenon has been experimentally observed \cite{14,15} and theoretically investigated \cite{11}. To date, owing to the appearance of a number of new optomechanical cavity systems, the phenomenon of several OMIT became the subject of theoretical research in atomic support-assisted optomechanical systems \cite{19,20}, hybrid piezoptomechanical cavity systems \cite{21} and multiple resonator optomechanical systems\cite{22}, etc. This complexity of light-matter interactions is also reflected in Fano resonance, initially observed in atomic systems \cite{40}. Fano resonance arises from the quantum interference of various transition amplitudes, producing minima in the absorption profile \cite{400}. Over the years, this phenomenon has been explored in a variety of physical systems, including photonic crystals \cite{038}, optomechanical systems \cite{41}, and coupled microresonators \cite{308}. Recently, experimental evidence of Fano-like asymmetric shapes has been reported in a hybrid cavity magnomechanical system \cite{36}. \\ 
	Recently, comparable to cavity optomechanics, the cavity magnomechanics system (CMM) \cite{39} has garnered increasing attention due to its numerous benefits over traditional systems. Magnetic materials typically have a high spin density and very low damping rate \cite{p,q}, which provides a great framework for the investigation of strong interactions between light and matter. Besides the advantages mentioned above, the magnon is highly capable of interacting with various quantum systems, like optical photons, superconducting qubits \cite{23},  microwaves \cite{25}, phonons \cite{26} and whispering gallery modes (WGM) \cite{27}. Because of these special characteristics, YIG-containing systems are used to study a variety of coherent phenomena similar to those of optomechanical systems, such as magnomechanical cooling and entanglement \cite{28}, magnomechanically induced transparency (MMIT) \cite{00}, bistability and non-reciprocity \cite{29,30}, fast-slow light engineering \cite{31,32}, ground-state cooling of magnomechanical resonators \cite{33,34}, and so on.\\

	Up to now, magnetically induced transparency (MIT) has been demonstrated successfully at room temperature, due to the quantum robustness of spin waves at this temperature \cite{35}. Due to the magnetostrictive interaction between magnon and phonon \cite{36}, magnetomechanically induced transparency (MMIT) and magnomechanical parametric amplification (MMPA) are observed in CMM \cite{36}. Tunable multiwindow Magnomechanically induced transparency regulated slow to fast light conversion have been proposed for the system, which consists of two ferromagnetic YIG spheres associated with a single microwave cavity mode \cite{00}. Furthermore, the generation of tripartite entanglement in a typical three-mode coupled cavity magnomechanical system has been studied \cite{37}. Recently, Qian $et$ $al.$ investigated a protocol for entangling two mechanical vibration modes in a cavity magnomechanical system \cite{38}. \\

	Motivated by the above works, we will study the magnomechanically induced transparency phenomenon, Fano resonance, and fast-slow light effect in a magnomechanical system in which two high-quality yttrium iron garnet (YIG) spheres and a degenerate OPA are contained within a microwave cavity, see Fig. \ref{a}. We study the effect of magnomechanical coupling and the gain of OPA on the absorption and dispersion spectra. In addition, we discuss the emergence of Fano resonance in the output field and investigate the appropriate system parameters for its observation. Furthermore, we discuss the phenomena of slow light propagation. We explain that the group delay depends on the tunability of the magnon-phonon coupling of the first YIG sphere and the gain of OPA.\\

	The structure of the paper is constructed as follows: In Section \ref{1}, we introduce the system, give the formula of its Hamiltonian and the appropriate quantum Langevin equations (QLEs) and calculate the output field. In Subsection \ref{2}, we discuss the magnomechanically induced transparency and analyze the influence of the optomechanical coupling strength $R_1$ and the gain of the OPA on the input spectrum. In Subsection \ref{3} we study the Fano resonance in the output field. In Section \ref{4}, we describe the probe field transmission and discuss the group delays for slow and fast light propagation. Concluding remarks close this paper.
	
	\section{MODEL}\label{1}
	
\begin{figure}[t]
		\centering
		\hskip-1.0cm\includegraphics[width=0.8\linewidth]{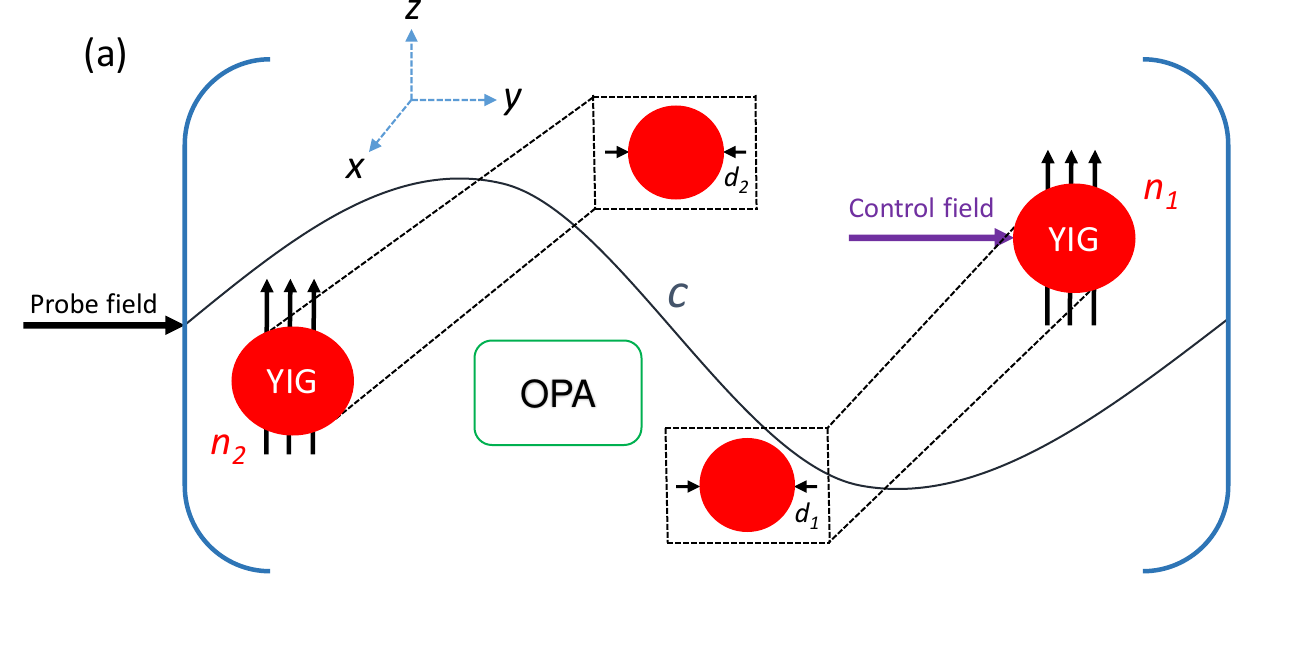}\\
		\hskip-1.0cm\includegraphics[width=0.8\linewidth]{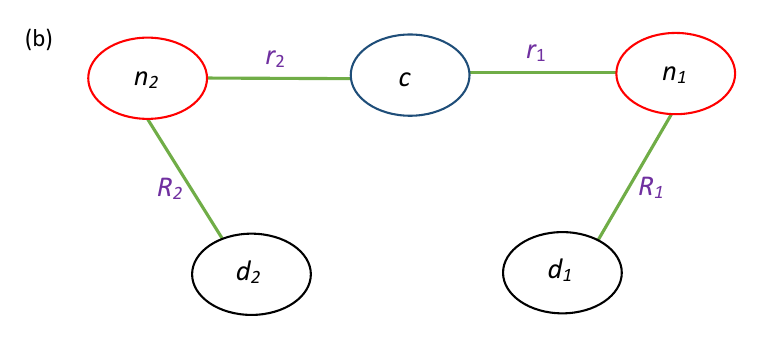}
		\caption{(a) Schematic representation of the system, comprising a magnomechanical cavity with two ferromagnetic yttrium iron garnet (YIG) spheres (shown in red) and a degenerate OPA (depicted by the green surface).  (b) The magnon mode $n_j$ (where $j=1,2$) interacts with the cavity mode $c$ with a coupling strength denoted by $r_j$, and it also couples to the mechanical mode $d_j$ with an effective magnomechanical coupling rate of $R_j$.} \label{a}
	\end{figure}

	A hybrid cavity magnomechanical system is depicted in Figure \ref{a}, featuring two ferromagnetic yttrium iron garnet (YIG) spheres and a degenerate optical parametric amplifier (OPA) within a microwave cavity. An externally applied magnetic field in the \textsf{z} direction excites the magnon modes in each sphere, which interact with the cavity field via magnetic dipole interaction. Additionally, phonon modes, which refer to the vibrations of the YIG spheres induced by the magnetostrictive force, facilitate magnon-phonon coupling within the spheres. The Hamiltonian that describes this system is given as
	\begin{equation}\label{a0}
		\begin{aligned}
			\mathcal{H} =\mathcal{ H}_{{free}}+\mathcal{H}_{{int}}+\mathcal{H}_{{drive}}+\mathcal{H}_{{OPA}},
		\end{aligned}
	\end{equation}	
	where the free Hamiltonian terms given by
	\begin{equation}
		\mathcal{H}_{{free}}=\hbar\omega_c c^{\dagger} c+\hbar\sum_{j=1,2}\left(\omega_{n_j} n_j^{\dagger} n_j+\frac{\omega_{d_j}}{2}\left(y_j^2+x_j^2\right)\right),
	\end{equation}
	the first and second terms describe the energy of the cavity and magnon modes, respectively, and $\omega_c$ is the frequency inside the cavity, and $\omega_{n_j}$ are the frequencies of each magnon. The magnon frequency is defined by the external bias magnetic field $\textsf{H}_j$ and the gyromagnetic ratio $\gamma_r$ i.e, $\omega_{n_j}=\gamma_r \textsf{H}_j$ with $\gamma_r/2\pi=28$ GHz/T \cite{28}. The operators $c(c^\dagger)$ and $n(n^\dagger)$ represent the annihilation and creation of the cavity and magnon modes, respectively; they satisfy the commutation relations $[c,c^\dagger]=1$; $[n_j,n_j^\dagger]=1$ with $j=1,2$. The third term represents the energy of two mechanical vibration modes with frequencies $\omega_{d_j}$, and $x_j$ and $y_j$ are respectively the dimensionless position and momentum of the phonon mode such that $[x,y]=i$. The interaction Hamiltonian terms are given by
	\begin{equation}
		\mathcal{H}_{{int}}=\hbar\sum_{j=1,2}\left(r_j\left(c n_j^{\dagger}+c^{\dagger} n_j\right)+R_{0 j} n_j^{\dagger} n_j x_j\right),	
	\end{equation}
	the first term denotes the energy of the interaction between the optical mode and magnon mode, with $r_j$ being the liner cavity-magnon coupling rate. The last term describes the energy of the interaction between the mechanical mode and the magnon mode, with $R_{0j}$ being the bare magnomechanical coupling rate. The drive Hamiltonian terms are given by
	\begin{equation}
		\mathcal{H}_{{drive}}=i\hbar \Omega_l\left(n_l^{\dagger} e^{-i \omega_{0 l} t}-\text{H.c}\right)+i\hbar\left( c^{\dagger}\epsilon_p e^{-i\omega_pt}-\text{H.c}\right),
	\end{equation}
	the first term represents the energy of the strong microwave field. This microwave field plays the role of a control field in our system. The Rabi frequency $\Omega_l=\frac{\sqrt{5}}{4}\gamma_r\sqrt{N_l}B_l$ $(l=1 $ or $ 2)$ is the strength of the driving field (see Appendix \ref{A}), with $N_l=\nu \mathcal{V}$ being the total number of spins in the $l$th YIG spheres with $\nu=4.22\times 10^{27}/\text{m}^3$ the spin density of the YIG and $\mathcal{V}$ is the volume of the sphere, and $B_l$ ($\omega_{0 l}$) represents the amplitude (frequency) of the drive magnetic field. The Rabi frequency $\Omega_l=\Omega_1\delta_{1l}$ $(l=1,2)$ means that the control field only applies to the magnon mode $n_1$. The second term denotes the energy of the optical driving, with $\omega_p$ and $\epsilon_p=\sqrt{2\kappa_cP_p/\hbar\omega_p}$ being, respectively, the frequency and the amplitude of the probe field, where $P_p$ is the power of the probe field and $\kappa_c$ is the cavity delay rate.

	The OPA Hamiltonian terms are given by
	\begin{equation}
		\mathcal{H}_{{OPA}}=i\lambda(e^{i\theta}{c^{\dagger}}^2e^{-i2\omega_{01}t}-e^{-i\theta}{c}^2e^{i2\omega_{01}t}),
	\end{equation}
	represents the coupling of the cavity field with the degenerate OPA; $\lambda$ and $\theta$ being, respectively, the gain and the phase of the field driving the OPA. In an OPA, a pump field at a frequency of $2\omega_{01}$ interacts with a second-order nonlinear optical crystal, resulting in the signal and idler both having the same frequency of $\omega_{01}$. In a frame rotating at the frequency $\omega_{01}$, the quantum Langevin equations describing the system dynamics can be written as
	\begin{equation}\label{a1}
		\begin{aligned}
			\dot{c}=  - & \left(i \Delta_c+\kappa_c\right) c -i r_1 n_1-i r_2 n_2+\epsilon_p e^{-i\delta t}\\
			& +2\lambda e^{i\theta}c^\dagger+\sqrt{2 \kappa_c} c^{i n}, \\
			\dot{n}_j  =- & \left(i \Delta_{n_j}+\kappa_{n_j}\right) n_j-i R_{0 j} n_j x_j-i r_j c \\
			& +\Omega_j+\sqrt{2 \kappa_{n_j}} n_j^{i n}, \\
			\dot{x}_j & =\omega_{d_j} y_j, \\
			\dot{y}_j & =-\omega_{d_j} x_j-\gamma_{d_j} y_j-R_{0 j} n_j^{\dagger} n_j+ {\zeta _j},
		\end{aligned}
	\end{equation}
	where $\delta=\omega_p-\omega_{01}$, $\Delta_c=\omega_c-\omega_{01}$ and $\Delta_{n_j}= \omega_{n_j}-\omega_{01}$ are the detunings. $\kappa_{n_j}$ ($\gamma_{d_j}$) denotes the dissipation rates of the magnon (mechanical) mode. $c^{i n}$ ($n_j^{i n}$) is the input noise operators affecting the cavity (magnon) modes. $\zeta _j$ is the hermitian Brownian noise operator acting on the mechanical mode. \\
The strong drive field leads to the excitation of large-amplitude magnon modes $|n_{js}|\gg 1$ ($j=1,2$), and cavity mode $|c_s|\gg 1$, through magnon-cavity coupling. In this case, the steady-state solutions for the first order in $\epsilon_p$, writes as $\langle \mathcal{X}\rangle=\mathcal{X}_{s}+\mathcal{X}_{-} e^{-i \delta t} +\mathcal{X}_{+}e^{i \delta t}$, where $\mathcal{X}= c, n_1, n_2, x_1, x_2, y_1,y_2$. Thus, the steady state solutions of the dynamic operator are given by 
	\begin{equation}
		c_{s}=\frac{-ir_1n_{1s}-ir_2n_{2s}+2\lambda e^{i\theta}c_s^*}{\kappa_c+i\Delta_c},
	\end{equation}
	\begin{equation}
		n_{1s}=\frac{-ir_1c_s+\Omega_1}{\kappa_{n_1}+i\tilde{\Delta}_{n_1}},
	\end{equation}
	\begin{equation}
		n_{2s}=\frac{-ir_2c_s}{\kappa_{n_2}+i\tilde{\Delta}_{n_2}},
	\end{equation}
	\begin{equation}
		x_{1s}=\frac{-R_{01}|n_{1s}|^2}{\omega_{d_1}},
	\end{equation}
	\begin{equation}
		x_{2s}=\frac{-R_{02}|n_{2s}|^2}{\omega_{d_2}},
	\end{equation}
	where $\bar{\Delta}_{n_1}=\Delta_{n_1}+R_{01}x_{1s}$ and $
	\bar{\Delta}_{n_2}=\Delta_{n_2}+R_{02}x_{2s}$. \\
	The solution for the cavity modes is given by (see Appendix \ref{Ap})
	\begin{equation}
		c_-=\frac{\eta\chi\varsigma\epsilon_p}{\alpha\eta\chi\varsigma+ir_1\sigma\chi\varsigma+ir_2\beta\eta\varsigma-2\varrho\eta\chi \lambda e^{i\theta}}
	\end{equation}
	According to the input-output standard relation for the cavity field: $\epsilon_{out}=\epsilon_{int}-2\kappa_c\langle c\rangle$ \cite{11,180}, the amplitude of the output field is given by 
	\begin{equation} \label{c}
		\epsilon_{out}=\frac{2\kappa_cc_-}{\epsilon_p}.
	\end{equation}
	The real part Re[$\epsilon_{out}$] and imaginary part Im[$\epsilon_{out}$] describes the absorption and dispersion spectrum of the output probe field, respectively.
	\section{RESULTS AND DISCUSSION } 
	\subsection{Magnomechanically induced transparency}\label{2}
	
	In this subsection, we numerically study the effect of the interaction between magnon $n_1$ and phonon $b_1$, as well as the degenerate OPA on the magnomechanically induced transparency in this hybrid cavity magnomechanical system. We use the effective parameters from a recent experiment in the cavity magnomechanical system, like \cite{39}: $\omega_c/2 \pi=10$ GHz, $\omega_{d}/2 \pi=\omega_{d_1}/2 \pi=\omega_{d_2}/2 \pi=10$ MHz, $\gamma_{d_1}/2 \pi=\gamma_{d_2}/2 \pi=100$ Hz, $\omega_{1,2}/2 \pi=10$ GHz, $\kappa_c/2 \pi=2.1$ MHz, $\kappa_{n_1}/2 \pi=\kappa_{n_2}/2 \pi=0.1$ MHz, $r_1/2\pi=r_2/2\pi=1.5$ MHz, $\Delta_c=\omega_{d_1}$, $\omega_{n_j}=\omega_{d_1}$ ($j=1,2$ ) and $\omega_{01}/2\pi=10$ GHz. The amplitude of the drive magnetic field $B_{1,2}=3.6\times 10^{-5}$ T, corresponding to the drive power $P_{1,2} = 7.6$ mW and the coupling $R_{0j}/2\pi=0.2$ Hz for a 250-$\mu$m-diameter YIG sphere. In this situation the number of spins $N \simeq 3.5 \times 10^{16}$ corresponds to $|\langle n_{js} \rangle| \simeq 1.1 \times 10^7$, leading to $\langle n_j^{\dag} n_j \rangle \simeq 1.2 \times 10^{14} \ll 5N=1.8 \times 10^{17}$, ($j=1,2$), which is well fulfilled.\\ 
	\begin{figure} [h!] 
		\begin{center}
			\includegraphics[scale=0.21]{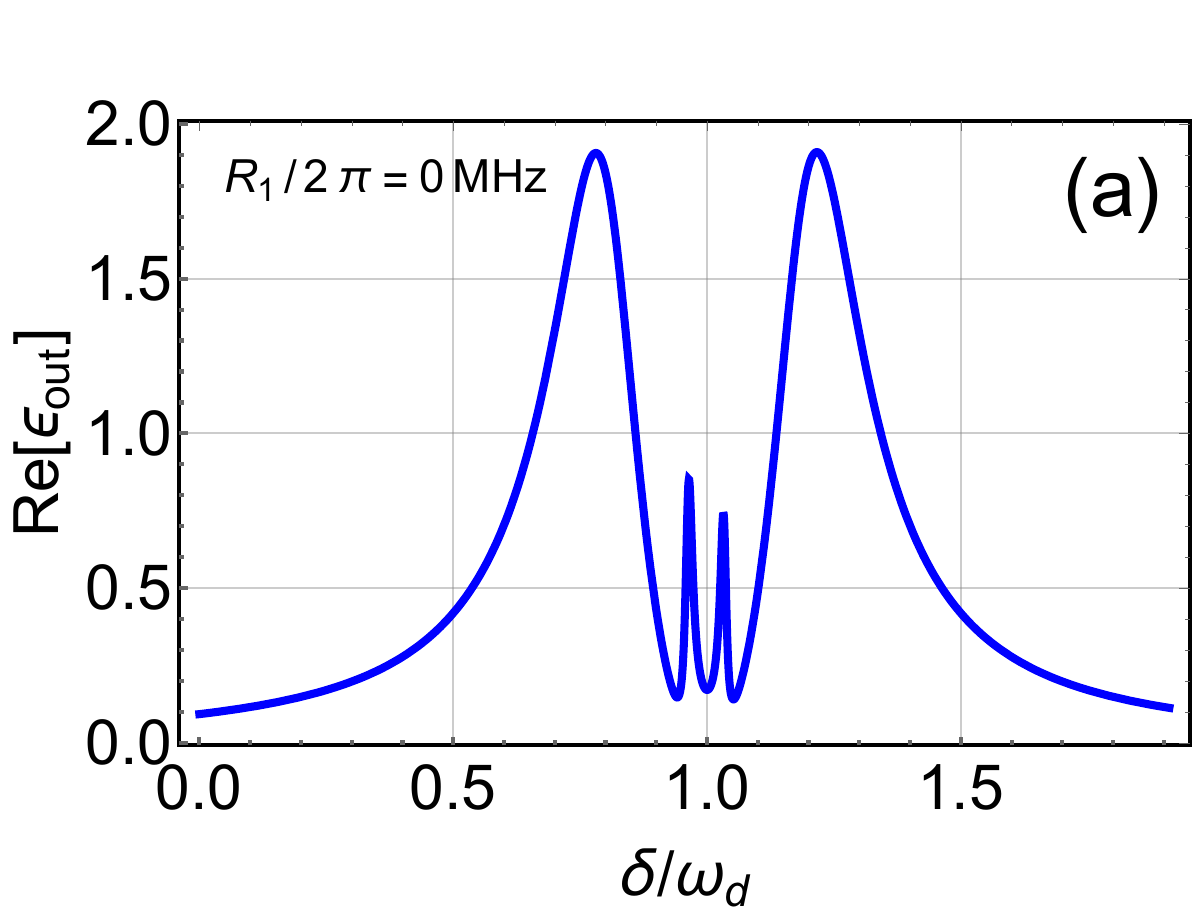}
			\includegraphics[scale=0.21]{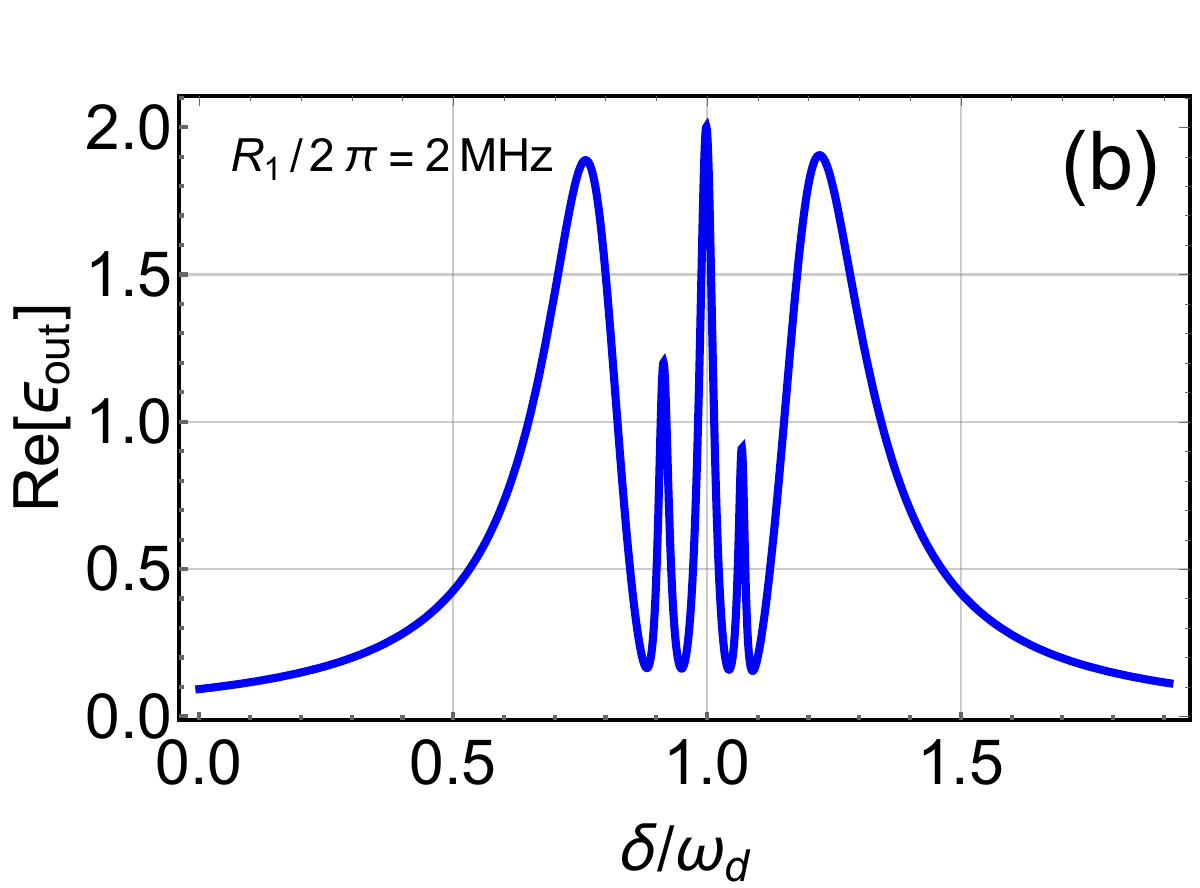}\\
			\includegraphics[scale=0.21]{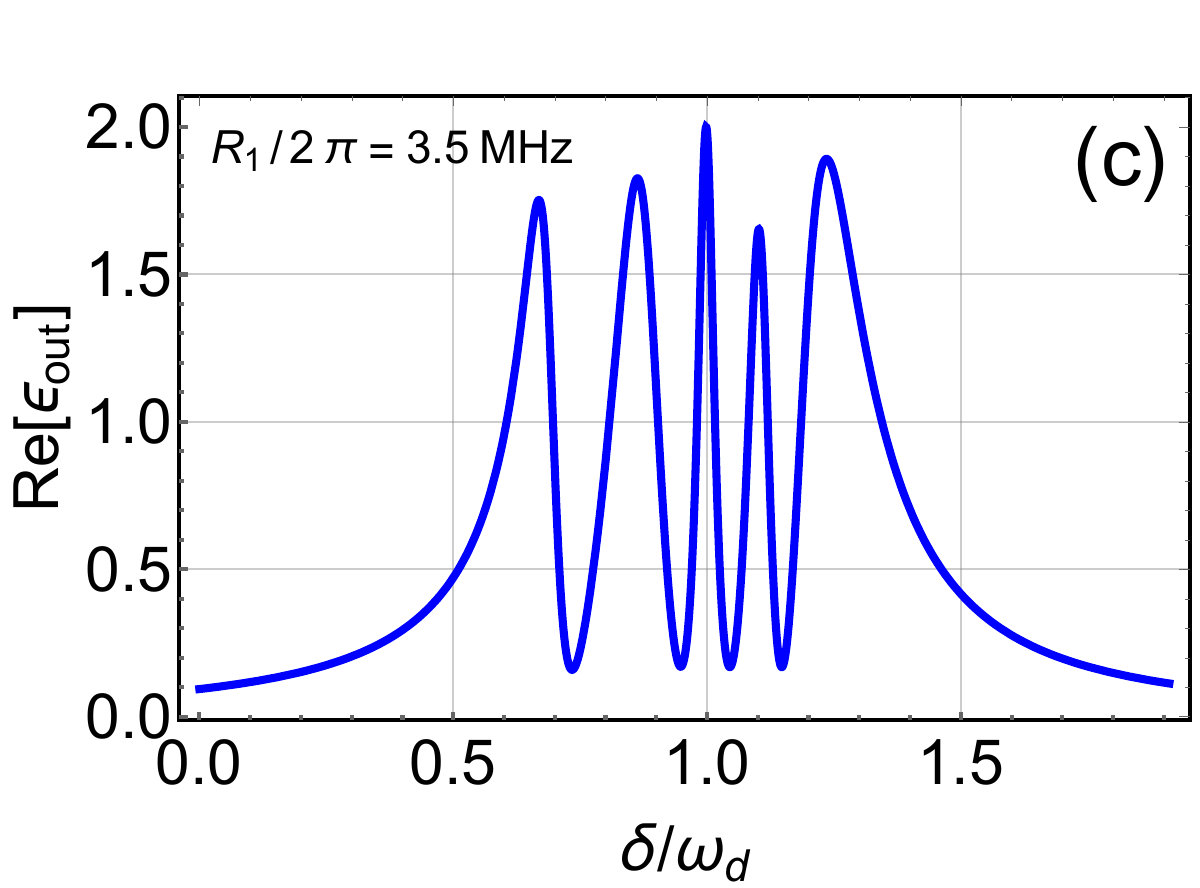}
			\includegraphics[scale=0.21]{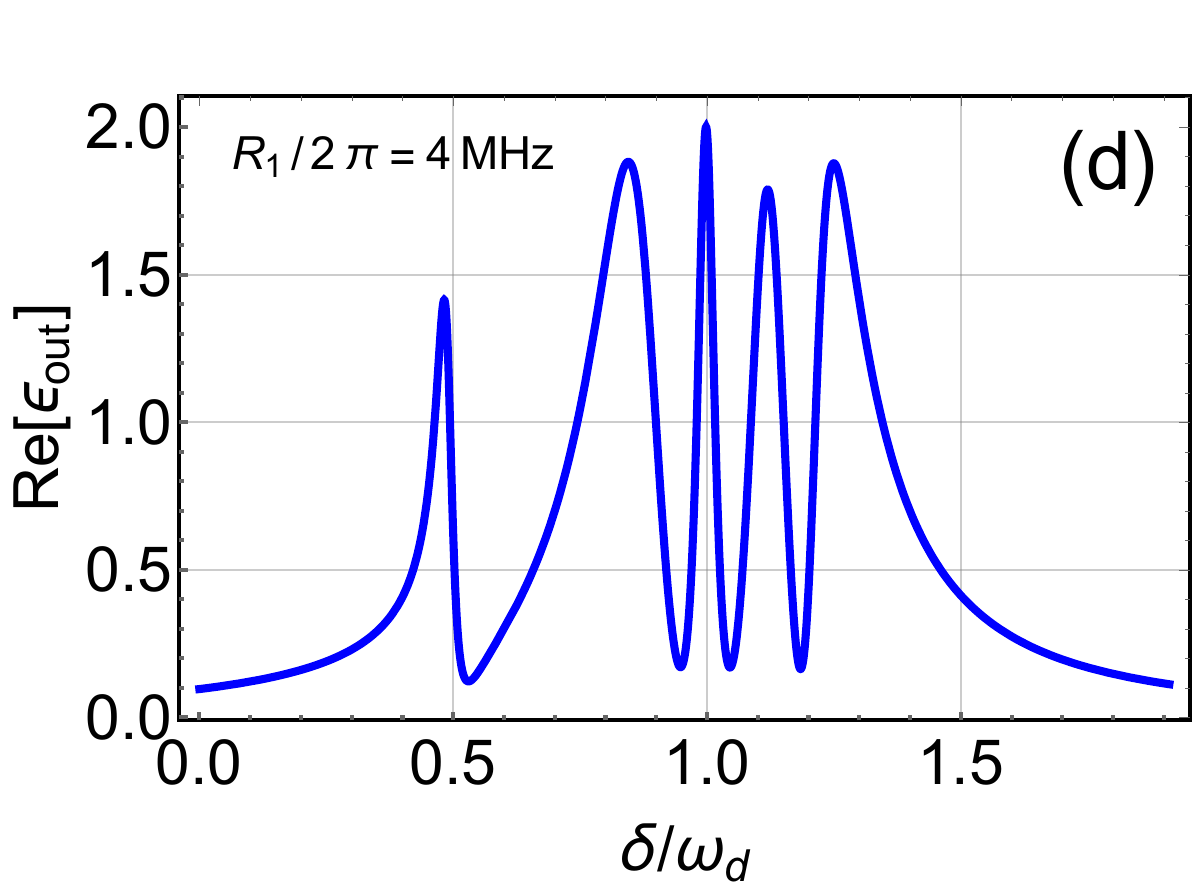}
			\caption{Plots of the real part of the output field Re[$\epsilon_{out}$] versus of $\delta/\omega_d$ for several values of the magnon-phonon coupling $R_1$ with $R_2/2\pi=1$ MHz and $\lambda=0$. (a) $R_1=0$ MHz, (b) $R_1/2\pi=2$ MHz, (c) $R_1/2\pi=3.5$ MHz, and (d) $R_1/2\pi=4$ MHz.} 
			\label{b}
		\end{center}
	\end{figure} 
	In Figure \ref{b}, we plot the Re[$\epsilon_{out}$] versus the normalized probe detuning $\delta/\omega_d$ for various values of magnomechanical coupling $R_1$. In Fig. \ref{b}(a), we consider the case where the interaction between the magnon $n_1$ and the phonon $d_1$ is neglected, i.e., $R_1$ is set to zero. Consequently, only the magnons $n_1$ and $n_2$ are coupled with the cavity, and the magnon $n_2$ is coupled with the phonon $d_2$. From this consideration, we can observe three transparency window dips, which were previously discussed in \cite{00}. However, for $R_1/2\pi=2$ MHz, five absorption peaks appear, separated by four MMITs, as depicted in Fig. \ref{b}(b). Therefore, two windows are associated with the magnomechanical interactions $R_1$ and $R_2$, and the other two are induced by magnon-photon couplings. By increasing the magnomechanical coupling $R_1$, the two small peaks increase in height and the widths of the transparency windows become broader. In Fig. \ref{b}(d), we see a shift to the left of the left transparency window when increasing the $R_1$.  \\ 
	\begin{figure} [h!] 
		\begin{center}
			\includegraphics[scale=0.21]{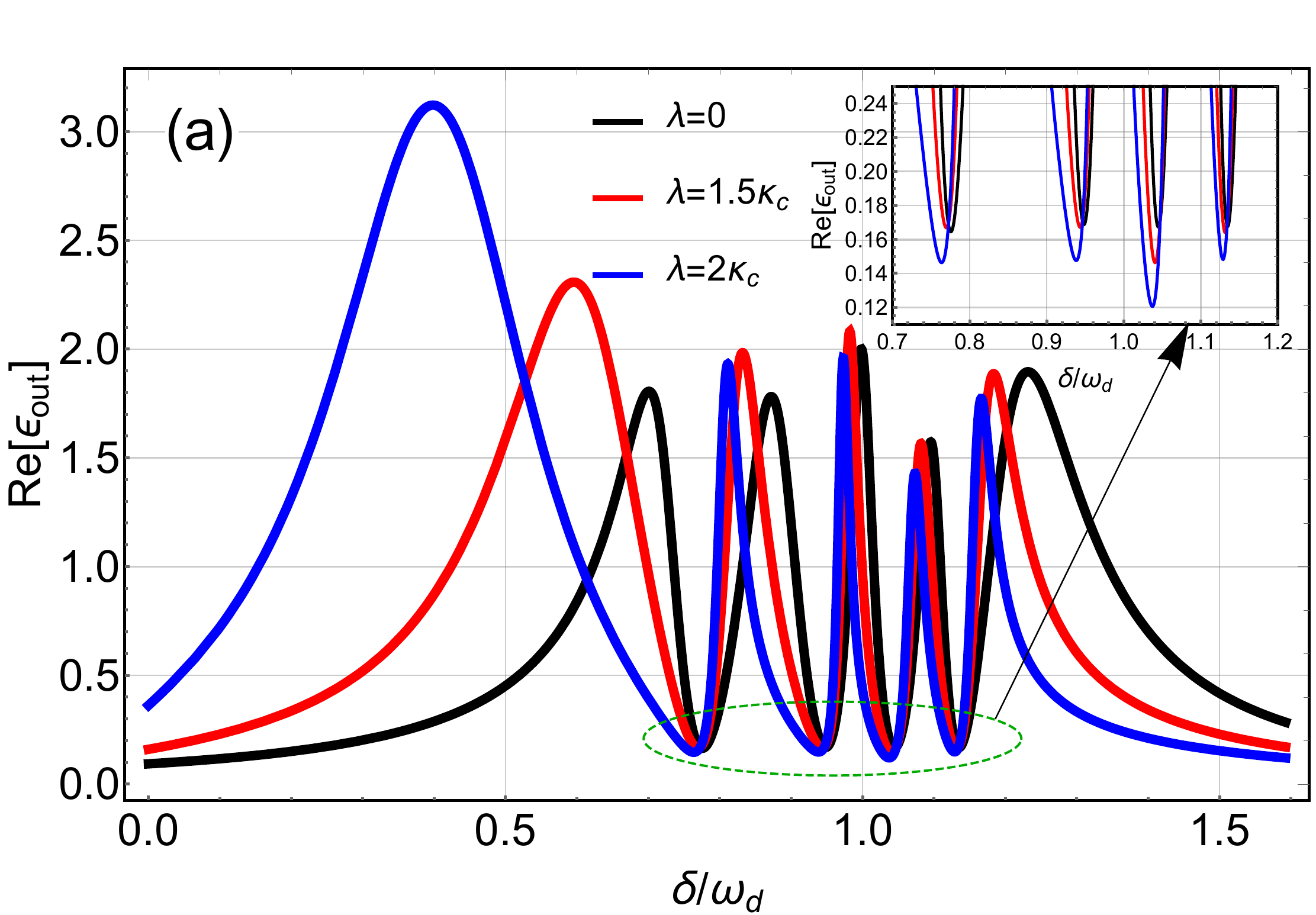}
			\includegraphics[scale=0.21]{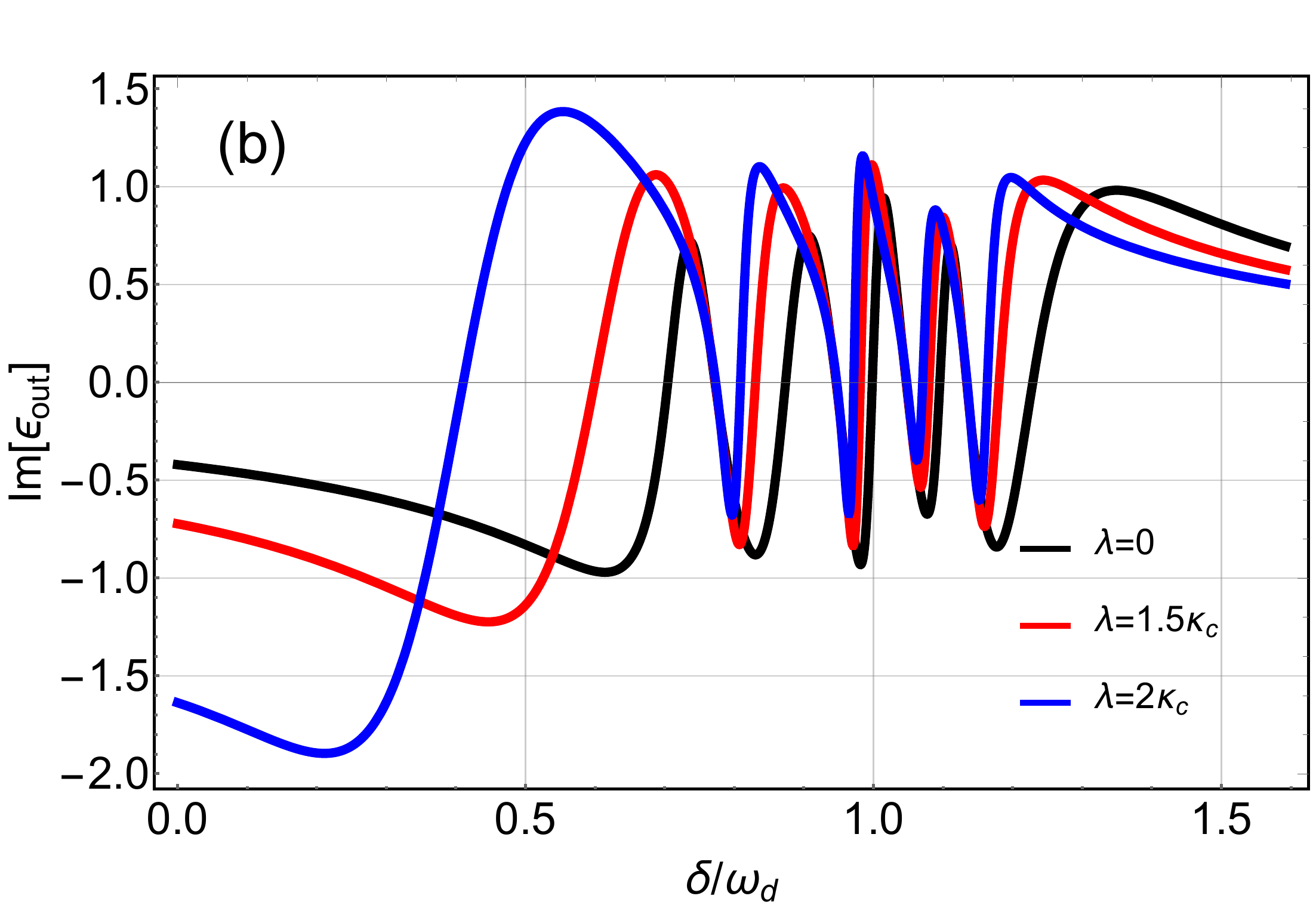}
			\caption{Plots of absorption Re[$\epsilon_{out}$] and dispersion Im[$\epsilon_{out}$] spectra of the output field versus of $\delta/\omega_d$ for several values of the gain $\lambda$ with $\theta=0$, $R_1/2\pi=1$ MHz, and $R_2/2\pi=3.5$ MHz.} \label{d}
		\end{center}
	\end{figure} 
	We plot in Fig. \ref{d}, the absorption Re[$\epsilon_{out}$] and the dispersion  Im[$\epsilon_{out}$] as a function of the normalized probe field detuning $\delta/\omega_d$ for different values of the gain $\lambda$ with $\theta=0$. In Fig. \ref{d}(a), we examine the case where $R_1/2\pi=1$ MHz, and $R_2/2\pi=3.5$ for different $\lambda$. We observe that as the gain of the degenerate OPA increases, the transparency window becomes wider and deeper. This suggests that the MMITs effect increases with the increase of $\lambda$. Consequently, the width of the transparent window can be regulated by adjusting the gain of the degenerate OPA. In Fig. \ref{d}(b) we plot the output field dispersion for various values of the gain $\lambda$ against of the normalized detuning $\delta/\omega_d$. The breadth of the transparency windows is clearly increasing with the gain of the degenerate OPA. 
	
\subsection{Fano resonances} \label{3}
	
In this subsection, we will discuss how Fano line shapes appear in the output spectrum as well as their physical mechanism. The Fano resonance exhibits a distinctly asymmetric shape, unlike the symmetric resonance curves observed in EIT, optomechanically induced transparency (OMIT) and magnomechanically induced transparency windows \cite{40,41}. Fano resonances have been detected in systems where EIT phenomena were previously reported, achieved through careful adjustment of system parameters \cite{42,43,44}. The unique asymmetric profile of Fano resonances observed in systems with optomechanical-like interactions is attributed to the non-resonant interactions. For instance, within a conventional optomechanical framework, the emergence of asymmetric Fano resonance shapes in the spectral distribution can occur when the anti-Stokes process does not align with the cavity's resonant frequency \cite{41,42}. 
	
\begin{figure} [h!] 
		\begin{center}
			\includegraphics[scale=0.21]{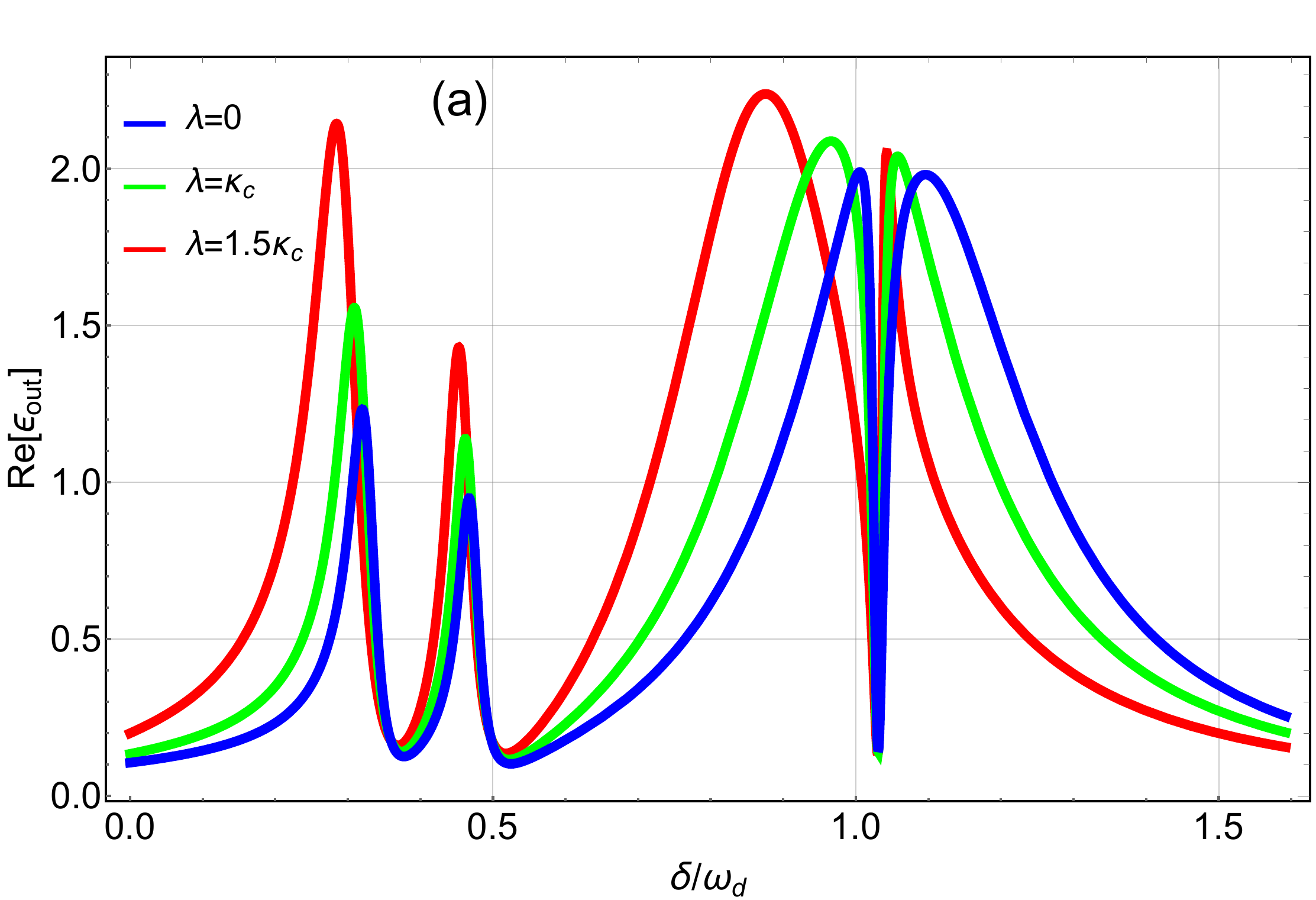}
			\includegraphics[scale=0.19]{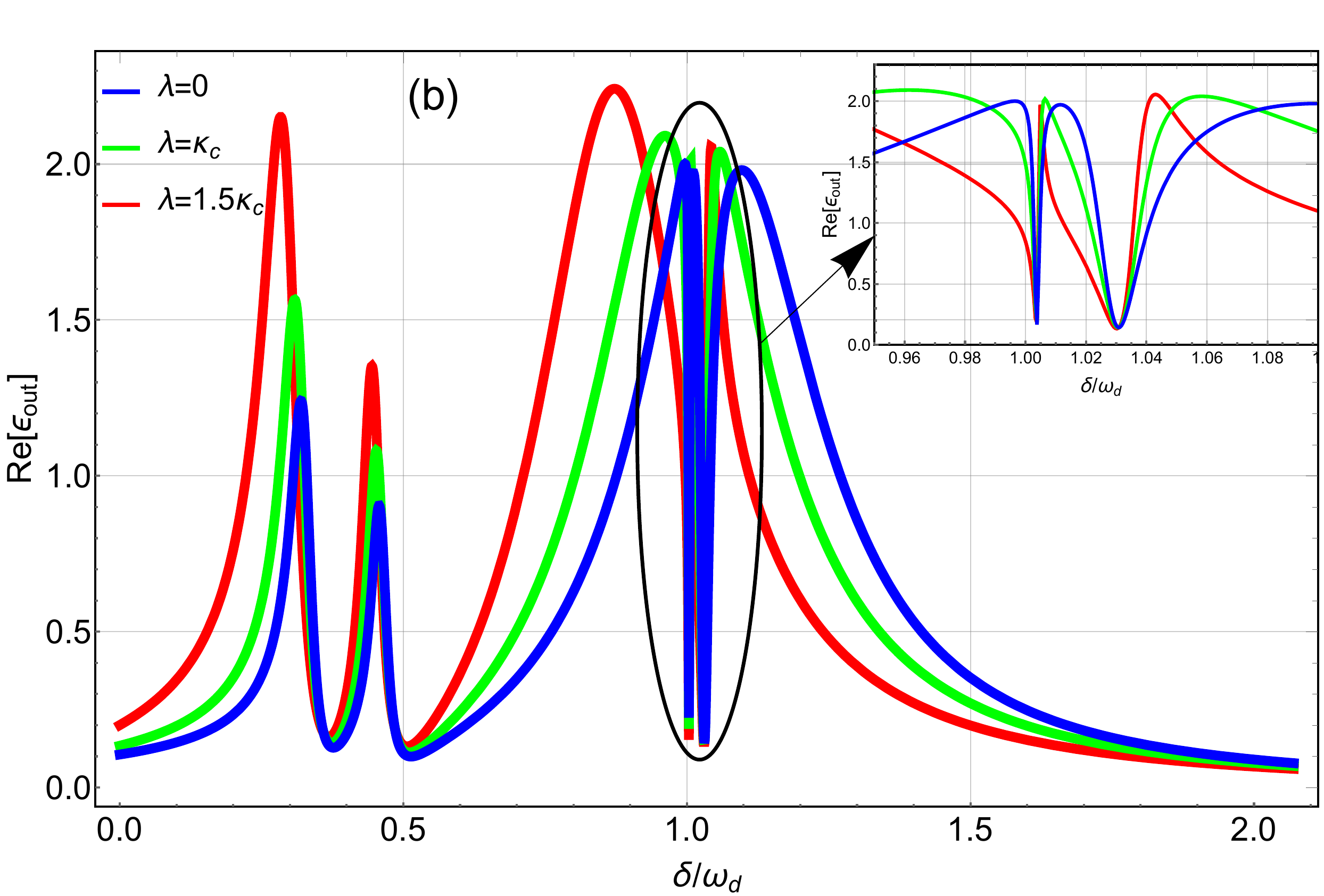}
			\includegraphics[scale=0.21]{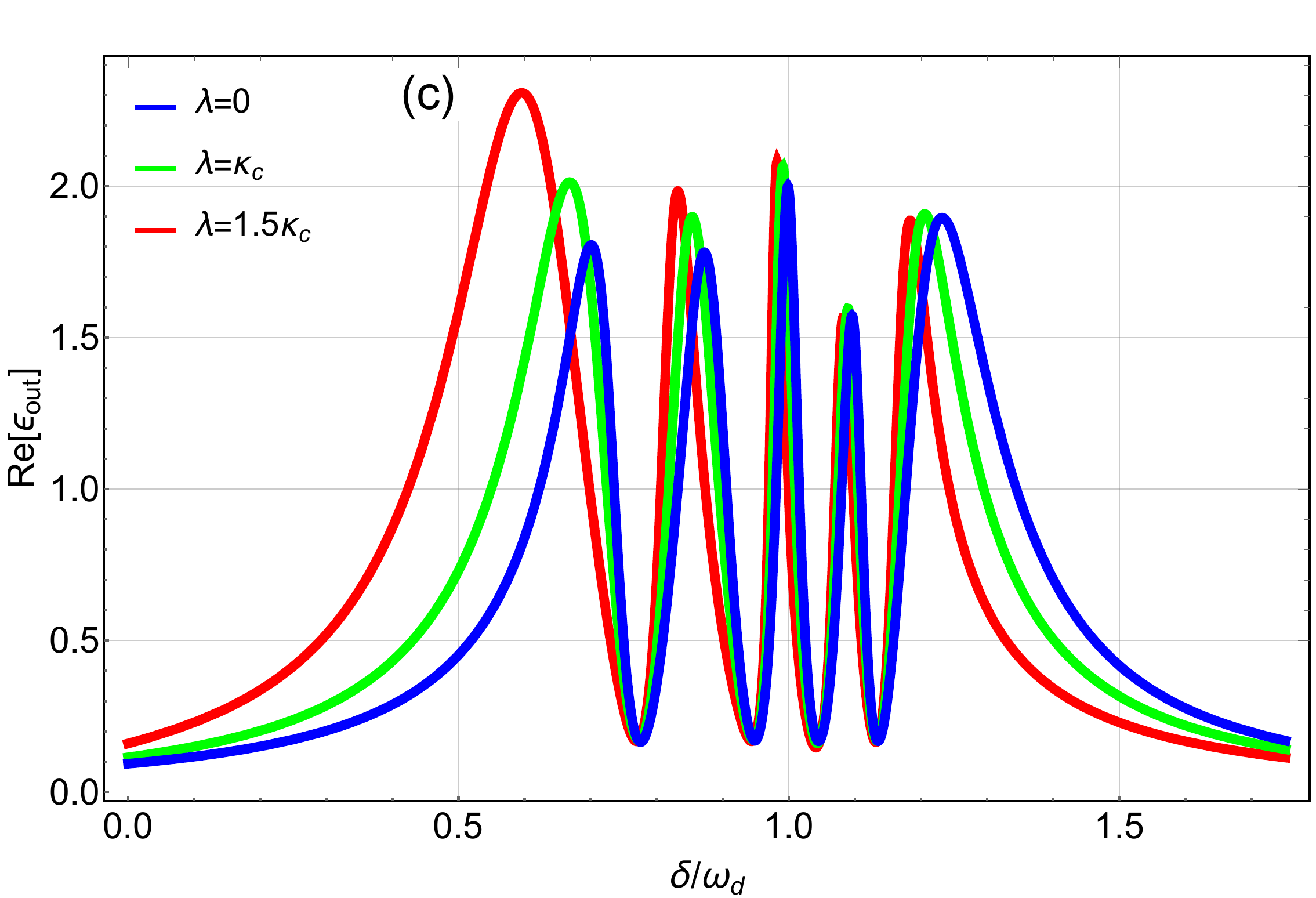}
			\caption{ Plot of Fano line shapes in the asymmetric absorption  Re[$\epsilon_{out}$] versus of the normalized detuning $\delta/\omega_d$ for several values of the gain $\lambda$. (a) $\Delta_{n_{1,2}}=0.5 \omega_{d}$, $r_{1}/2\pi=r_{2}/2\pi=1.5$ MHz, $R_{2}/2\pi=3.5$ MHz, and $R_{1}=0$, and (b) $\Delta_{n_{1,2}}=0.5 \omega_{d}$, $r_{1}/2\pi=r_{2}/2\pi=1.5$ MHz, $R_{2}/2\pi=3.5$ MHz, and $R_{1}/2\pi=1$ MHz. (c) $\Delta_{n_{1,2}}=\omega_{d}$, $r_{1}/2\pi=r_{2}/2\pi=1.5$ MHz, $R_{2}/2\pi=3.5$ MHz, and $R_{1}/2\pi=1$ MHz.} \label{e}
		\end{center}
	\end{figure}

The asymmetric Fano shapes are observable in Figs. \ref{e}(a) and \ref{e}(b) for various nonresonant cases, where the real part of the output field is plotted against the normalized detuning $\delta/\omega_d$. In Fig. \ref{e}(a), the coupling between the magnon mode $n_1$ and the mechanical mode $d_1$ is set to zero.  Therefore, only magnons $n_1$ and $n_2$ are coupled with the cavity, and magnon $n_2$ is coupled with phonon $d_2$.  Owing to the presence of non-resonant ($\Delta_{n_{1,2}}=0.5 \omega_{d_1}$), we find a triple Fano resonance in the output field, as previously discussed in \cite{00}. Indeed, in the presence of all four couplings and $\Delta_{n_{1,2}}=0.5 \omega_{d_1}$ the triple Fano resonance transforms into a four Fano profile, as seen in Fig. \ref{e}(b). This occurs because the cavity field can be constructed through four coherent pathways provided by the four coupled systems (cavity, two magnons, and phonon modes), which are capable of interfering with each other. From Figs \ref{e}(a) and \ref{e}(b), it is clear that as the nonlinear gain of the degenerate OPA increases, the Fano resonance becomes wider. In Fig. \ref{e}(c), Fano resonances disappear when considering the resonant case $\Delta_{n_1}=\Delta_{n_1}=\omega_{d}$. 
	
\section{SLOW AND FAST LIGHT} \label{4}
	
This section explores how the transmission of a probe light field can be manipulated to exhibit both slow and fast light behavior within a magnomechanical system. According to Eq. \ref{c}, the rescaled transmission field associated with the probe field can be written as follows
	\begin{equation}
		\mathbb{T}=\frac{\epsilon_p-2\kappa_cc_-}{\epsilon_p}.
	\end{equation}
	
	Within the narrow transparency window, the rapid variation of the phase, denoted by $\Phi=\text{Arg}[\mathbb{T}]$, significantly impacts the group delay, expressed as
	
	\begin{equation}
		\tau  = \frac{{\partial \Phi }}{{\partial {\omega _p}}} = {\mathop{\rm Im}\nolimits} \left[ \frac{1}{\mathbb{T}}\frac{{\partial \mathbb{T}}}{{\partial {\omega _p}}}\right]. 
	\end{equation}
	A positive group delay ($\tau>0$) signifies slow light, while a negative delay ($\tau<0$) indicates fast light propagation.

\begin{figure} [h!] 
		\begin{center}
			\includegraphics[scale=0.4]{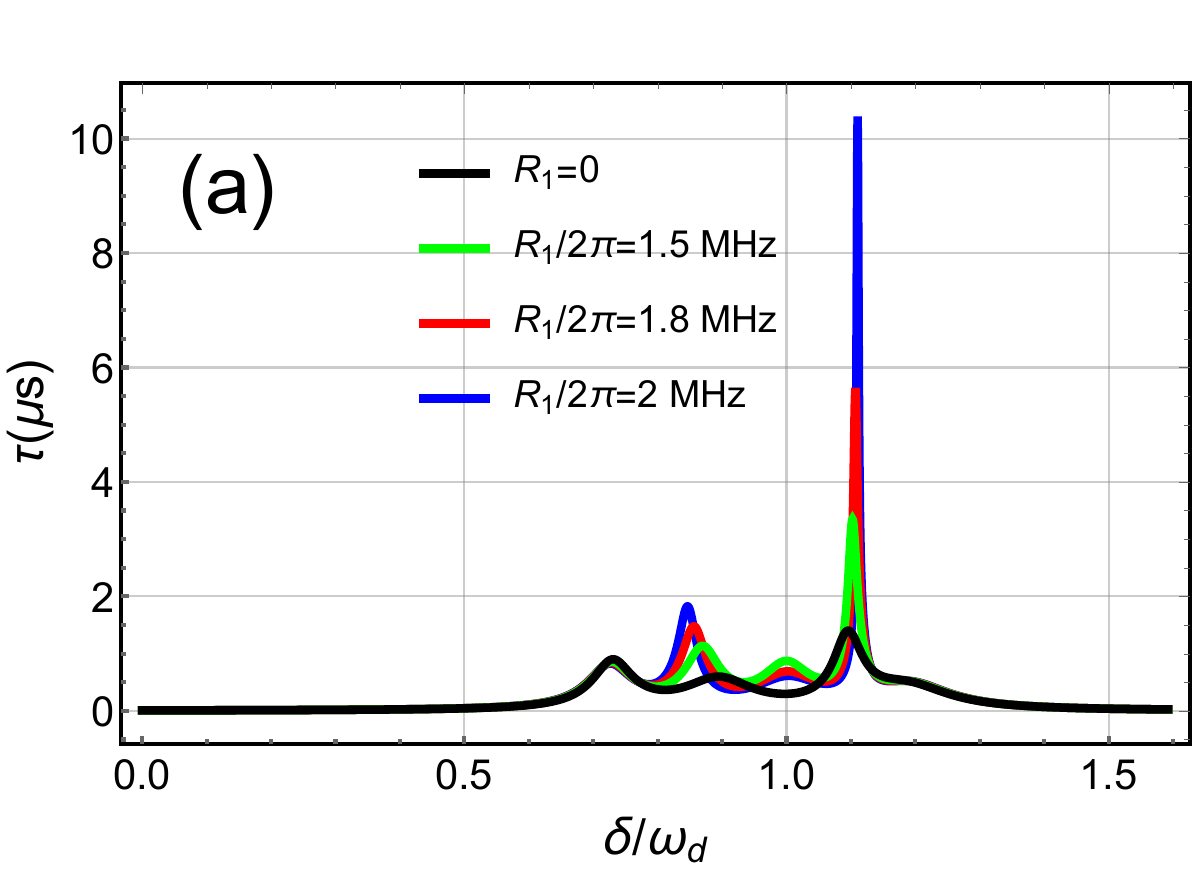}
			\includegraphics[scale=0.4]{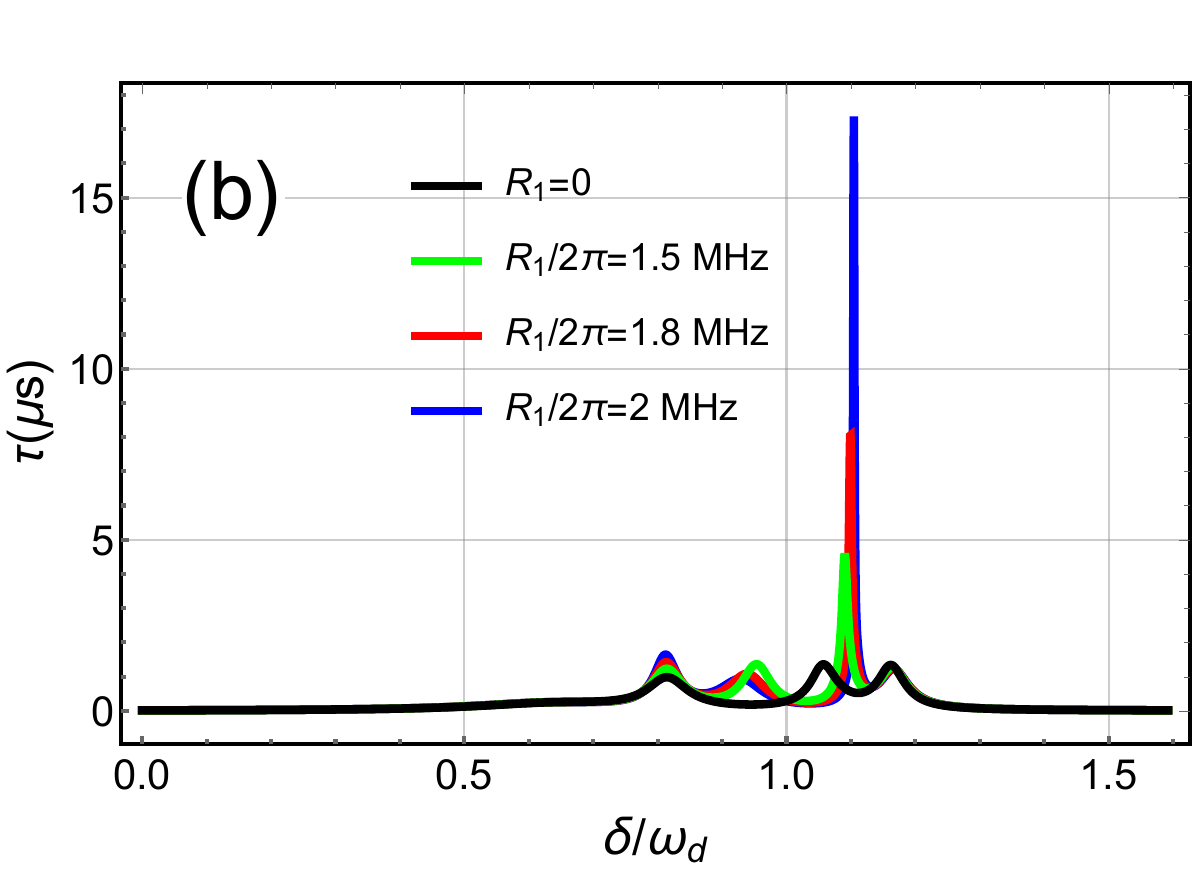}
			\caption{Group delay $\tau$ versus of normalized detuning $\delta/\omega_{d}$ for several values of phonon-magnon coupling $R_1$ with $R_2/2\pi=3.5$ MHz. (a) $\lambda=0$ and (b) $\lambda=1.5\kappa_c$.} \label{f}
		\end{center}
	\end{figure}

	In Figure \ref{f} we plot the group delay $\tau$ of the output field at the probe field frequency versus the normalized detuning $\delta/\omega_{d}$ for different values of magnomechanical coupling $R_1$ for a fixed value of all other parameters. We remark that the group delay of the output field is always positive and increases with increasing the magnomechanical coupling $R_1$. As a result, we can only observe the phenomenon of slow light. The ability to control the group delay arises from the interaction between the probe field and the anti-Stokes field through quantum interference. This implies the possibility of dynamically manipulating the group delay to achieve controllable slow light by adjusting the coupling strength, $R_1$, between phonons and magnons. Furthermore, the comparison between Figs. \ref{f}(a) and \ref{f}(b) shows that the increase in the gain of the degenerate OPA leads to an increase in group delay for the different values of the coupling strength $R_1$. For example the group delay value $\tau=10.38 \mu \text{s}$ when $\lambda=0$ and $R_1/2\pi=2$ MHz while $\tau=17.36 \mu \text{s}$ when $\lambda=1.5\kappa_c$ and $R_1/2\pi=2$ MHz.
	
	\begin{figure} [h!] 
		\begin{center}
			\includegraphics[scale=0.25]{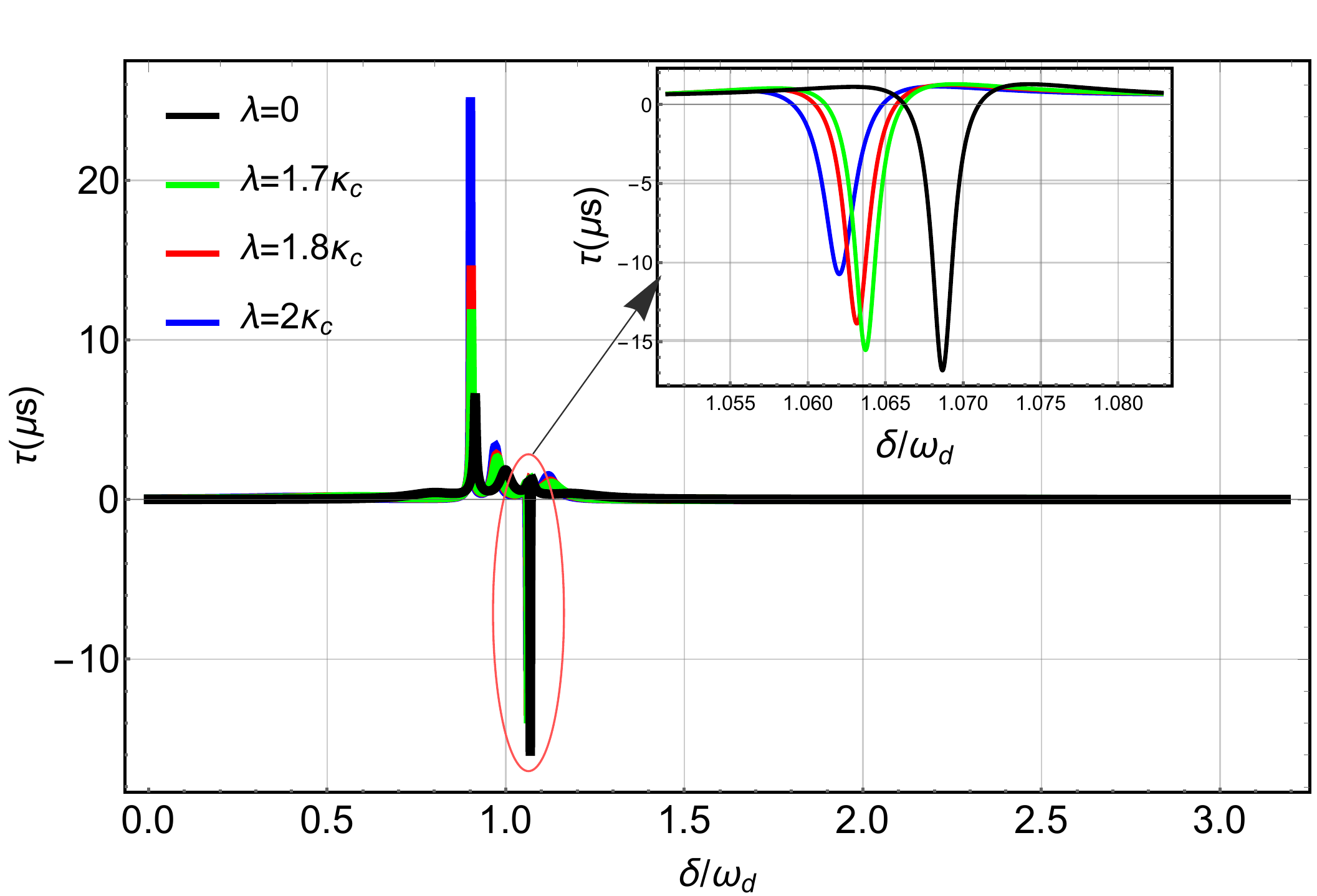}
			\caption{Group delay $\tau$ versus of normalized detuning $\delta/\omega_{d}$ for different values of the gain $\lambda$ with $R_1/2\pi=2$ MHz and $R_2/2\pi=1$ MHz.} \label{g}
		\end{center}
	\end{figure} 
	In Figure \ref{g}, we present the group delay $\tau$ of the output field at the probe field frequency versus the normalized detuning $\delta/\omega_{d}$ for various values of the gain $\lambda$ with $R_1/2\pi=2$ MHz and $R_2/2\pi=1$ MHz. Setting the optical parametric amplification (OPA) to zero ($\lambda=0$), we observe a clear signature of slow light in the output probe field. At a normalized detuning of $\delta/\omega_{d}\approx 0.91 $, the group delay of the output field reaches a positive value of $\tau\approx 6.61$ $\mu \text{s}$. In contrast, when the normalized detuning reaches $\delta/\omega_{d}\approx 1.06 $, the output field exhibits a negative group delay of $\tau\approx -16.87$ $\mu \text{s}$. This signifies the fast light effect experienced by the probe field. When a degenerate optical parametric amplifier (OPA) is introduced into the system, with coupling strengths $\lambda$ set to $1.7\kappa_c$, $1.8\kappa_c$, and $2\kappa_c$, we observe a significant impact on the magnomechanical system. This effect manifests as a pronounced enhancement in the positive group delay and a suppression of the negative group delay experienced by the output field. So we remark that when the degenerate OPA is placed inside the coupled system of two YIG spheres, the quantum interference effect between the anti-Stokes field and the probe field generated by the two magnons is directly related to the normalized detuning $\delta/\omega_{d}$ by the gain of the degenerate OPA.\\ 
	
	In the above results, the influence of the Kerr nonlinear term $K n_j^{\dag}n_j n_j^{\dag} n_j$ ($j=1,2$) on the Hamiltonian can be considered negligible~\cite{29,You16}. This term, with $K$ representing the Kerr coefficient that scales inversely with the system's volume, describes nonlinearities in the system. This unwanted Kerr nonlinear term may cause by the strong magnon pump. The Kerr coefficient, which characterizes the strength of a nonlinear effect in the material, is estimated to be around $10^{-10}/2\pi$ Hz for a YIG sphere with a diameter of 1 mm, as used in previous studies \cite{29,You16}. Because the coefficient scales inversely with the volume of the sphere, a smaller diameter of 250 micrometers would result in a roughly six times higher Kerr coefficient, approximately $6.4 \times 10^{-9}/2\pi$ Hz. Considering the values used in the figures, a term accounting for nonlinearities in the system, $K|\langle n_j \rangle|^3$ ($j=1,2$), is estimated to be around $5.7 \times 10^{13}$ Hz. This value is significantly smaller much less than compared to another parameter $\Omega_l$, which is $\Omega_l \geq  2.23\times 10^{14}$ Hz ($l$ is 1 or 2). This large difference suggests that the nonlinear effects can be considered negligible.
	
\section{Conclusion}

In this work, we investigate the interplay of multiple MMIT, Fano resonances, and slow-to-fast light effects in a cavity magnomechanical system comprising two YIG spheres and a degenerate OPA. We observe MMIT and MIT in the absorption spectrum of a weak probe field interacting with a strong control field. This is attributed to the phonon-magnon and photon-magnon interactions, where phonon-magnon coupling can further enhance MMIT and MIT. We then explore the influence of the OPA on the absorption and dispersion spectra. We find that optimizing the gain of the degenerate OPA can achieve a better transparency effect. Furthermore, we explain the emergence of Fano resonances in the probe field's output spectrum, arising from the presence of anti-Stokes processes within the system. Our study demonstrates the control of light speed by manipulating the coupling strengths between the phonon and magnon modes of the first YIG sphere, in conjunction with adjusting the OPA gain, enabling conversion between slow and fast light effects.
	\appendix
	\renewcommand{\thesection}{ \Alph{section}}
	\section{The Rabi frequency of the magnon $n_1$} \label{A}
	
	The Rabi frequency $\Omega_l$ represents the strength of the coupling between the driving magnetic field and the magnon mode. We proceed to derive its explicit expression below. This derivation is crucial as it will later enable us to determine the effective magnomechanical coupling rate and assess the accuracy of our linearized model.\\
	The equation $H = - \gamma_r \vec{s} \cdot \vec{B}$, where $\vec{s}$ is the spin angular momentum, is the standard form for the Hamiltonian of a spin $\vec{s}$ in a magnetic field $\vec{B}$. The interaction energy is determined by the gyromagnetic ratio $\gamma_r$, the spin, and the magnetic field. Because the YIG sphere contains a vast number of individual spins (represented by $\vec{s}$), it's more convenient to consider their combined effect. We define a new quantity called the collective spin angular momentum (represented by $\vec{S}$). This is obtained by summing the spin angular momentum of each individual spin within the sphere, as $\vec{S}=\sum \vec{s}$. When this collective spin interacts with an external, driving magnetic field. This field is oriented along the $y$-axis, has a constant amplitude of $B_{l}$, and oscillates at a specific frequency, $\omega_{0l}$. The Hamiltonian for this interaction $H_{d_l}$ is writes as $H_{d_l} = - \gamma_r \vec{S}_l \cdot \vec{B}_l = - \gamma_r S_{y_l} B_{l} \cos (\omega_{0l} t)$, where $\vec{S}_l =(S_{x_l}, S_{y_l}, S_{z_l})$. It is possible to express $H_d$ using the raising and lowering operators $S_l^{\pm}$, where $S_l^{\pm} = S_{x_l} \pm i S_{y_l}$, i.e., $H_{d_l} = i \frac{\gamma_r B_{l}}{4} (S_l^+ -S_l^-) (e^{i \omega_{0l} t} + e^{-i \omega_{0l} t})$. Through the Holstein-Primakoff transformation, $S_l^+ = \hbar  n_l \sqrt{2Ns - n_l^{\dag} n_l}\,$ and $S_l^- = \hbar \, n_l^{\dag}\! \sqrt{2Ns - n_l^{\dag} n_l}$, the collective spin operators $S_l^{\pm}$ are connected to the bosonic annihilation and creation operators of the magnon mode, $n_l$ and $n_l^{\dag}$, where the spin number of the ground state Fe$^{3+}$ ion in YIG is $s=\frac{5}{2}$ and $N$ is the total number of spins. $S_l^+ \approx \hbar \sqrt{5N}\, n_l$ and $S_l^- \approx \hbar \sqrt{5N}\, n_l^{\dag} $ represent the approximate values of the transformations described above for the low-lying excitations, $\langle n_l^{\dag} n_l \rangle \ll 2Ns$. This leads to the Hamiltonian 
	
	\begin{equation}
		\begin{split}
			H_{d_l} &= i \hbar \frac{\sqrt{5}}{4} \gamma_r \! \sqrt{N} B_{l} \, (n_l - n_l^{\dag}) (e^{i \omega_{0l} t} + e^{-i \omega_{0l} t})  \\
			&\approx  i \hbar \Omega_l \, (n_l e^{i \omega_{0_l} t}  -  n_l^{\dag}  e^{-i \omega_{0_l} t}), \quad l=\text{1 or 2},
		\end{split}
	\end{equation}
	with $\Omega_l =\frac{\sqrt{5}}{4} \gamma_r \! \sqrt{N} B_l$, $\gamma_r/2\pi= 28$ GHz/T, and $N=\nu \mathcal{V}$, where $\mathcal{V}$ represents the sphere's volume and $\nu=4.22 \times 10^{27}$ m$^{-3}$ represents the YIG's spin density.
	
\section{Derivation of $c_-$} \label{Ap} 
	
	$$\alpha=\kappa_c+i(\Delta_c-\delta),\, \alpha_1=\kappa_c-i(\Delta_c+\delta) ,\, \alpha_2=\kappa_{n_1}+i(\bar{\Delta}_{n_1}-\delta),$$
	$$\alpha_3=\kappa_{n_1}-i(\bar{\Delta}_{n_1}+\delta), \quad \alpha_4=\kappa_{n_2}+i(\bar{\Delta}_{n_2}-\delta),\quad $$
	$$\alpha_5=\kappa_{n_2}-i(\bar{\Delta}_{n_2}+\delta), \quad \alpha_6=\omega_{d_1}-\frac{\delta}{\omega_{d_1}}\left(i\gamma_{d_1}+\delta\right),$$$$\alpha_8=\omega_{d_2}-\frac{\delta}{\omega_{d_2}}\left(i\gamma_{d_2}+\delta\right), \quad \mathcal{A}=1+\frac{R_{22}^2}{i\alpha_4\alpha_8},$$  $$\mathcal{B}=1-\frac{R_{22}^2}{i\alpha_5\alpha_8\mathcal{A}},\quad \mathcal{C}=\frac{r_2G_{22}^2}{\alpha_4\alpha_5\alpha_8\mathcal{A}}, \quad \mathcal{D}=1+\frac{r_2^2}{\alpha_1\alpha_5\mathcal{B}}, $$ 
	$$  \mathcal{E}=\frac{ir_2\mathcal{C}}{\alpha_1\mathcal{B}}+\frac{2\lambda e^{-i\theta}}{\alpha_1}, \quad \mathcal{F}=1+\frac{r_1^2}{\alpha_1\alpha_3\mathcal{D}}-\frac{R_{11}^2}{i\alpha_3\alpha_6},$$
	$$ \mathcal{G}=\frac{ir_1\mathcal{E}}{\alpha_3\mathcal{D}}, \quad \mathcal{K}=\frac{R_{11}^2}{i\alpha_3\alpha_6}, \quad\eta=1+\frac{R_{11}^2}{i\alpha_2\alpha_6}+\frac{\mathcal{K}R_{11}^2}{i\alpha_2\alpha_6\mathcal{F}},$$  $$ \quad \sigma=\frac{\mathcal{G}R_{11}^2}{i\alpha_2\alpha_6\mathcal{F}}-\frac{ir_1}{\alpha_2}, \quad \mathcal{L}=1+\frac{R_{11}^2}{i\alpha_2\alpha_6},$$
	$$  \mathcal{M}=1-\frac{R_{11}^2}{i\alpha_3\alpha_6\mathcal{L}}, \quad \mathcal{N}=\frac{r_1R_{11}^2}{\alpha_2\alpha_3\alpha_6\mathcal{L}}, \quad \mathcal{O}=1+\frac{r_1^2}{\alpha_1\alpha_3\mathcal{M}},$$ $$\mathcal{P}=\frac{ir_1\mathcal{N}}{\alpha_1\mathcal{M}}+\frac{2\lambda e^{-i\theta}}{\alpha_1}, \quad \mathcal{Q}=1+\frac{r_2^2}{\alpha_1\alpha_5\mathcal{O}}-\frac{R_{22}^2}{i\alpha_5\alpha_8}, $$
	$$\mathcal{R}=\frac{ir_2F}{\alpha_5\mathcal{O}}, \quad \mathcal{U}=\frac{R_{22}^2}{i\alpha_5\alpha_8}, \quad \chi=1+\frac{R_{22}^2}{i\alpha_4\alpha_8}+\frac{\mathcal{U}R_{22}^2}{i\alpha_4\alpha_8\mathcal{Q}}, $$
	$$\beta=\frac{\mathcal{R}R_{22}^2}{i\alpha_4\alpha_8\mathcal{Q}}-\frac{ir_2}{\alpha_4}, \quad \varsigma=1+\frac{r_1^2}{\alpha_1\alpha_3\mathcal{M}}+\frac{r_2^2}{\alpha_1\alpha_5\mathcal{B}},$$
	$$\varrho=\frac{ir_1\mathcal{N}}{\alpha_1\mathcal{M}}+\frac{ir_2\mathcal{C}}{\alpha_1\mathcal{B}}+\frac{2\lambda e^{-i\theta}}{\alpha_1}.$$ \\
	Here, $R_{11}=\frac{R_1}{\sqrt{2}}$ and $R_{22}=\frac{R_2}{\sqrt{2}}$, with $R_j=i\sqrt{2}R_{0j}n_{js}$ representing the effective magnomechanical coupling rate, where $|\bar{\Delta}_{n_1}|,|\bar{\Delta}_{n_2}|,\left|\Delta_c\right| \gg \kappa_c, \kappa_{n_1},\kappa_{n_2}$.\\
	
\section*{Data availability statement}
No Data associated in the manuscript


\begin{thebibliography}{99}	
		\bibitem{1} K. J. Boller, A. Imamoğlu and S. E. Harris, Phys. Rev. Lett. {\bf 66}, 2593 (1991).
		\bibitem{2} M. Fleischhauer, A. Imamoglu and J. P. Marangos, Rev. Mod. Phys. {\bf 77}, 633-673 (2005).
		\bibitem{3} A. A. Abdumalikov Jr, O. Astafiev, A. M. Zagoskin, Y. A. Pashkin, Y. Nakamura and J. S. Tsai, Phys. Rev. Lett. {\bf 104}, 193601 (2010).
		\bibitem{4} P. Rabl, Phys. Rev. Lett. {\bf 107}, 063601 (2011).
		\bibitem{5} A. Nunnenkamp, K. Børkje and S. M. Girvin, Phys. Rev. Lett. {\bf 107}, 063602 (2011).
		\bibitem{6} Setodeh Kheirabady, M., E. Ghasemian, and M. K. Tavassoly. Annalen der Physik, {\bf 535.6}, 2300024 (2023).
		\bibitem{7} M. Amazioug, D. Dutykh, B. Teklu and M. Asjad, Ann. Phys. {\bf 536}, 2300357 (2023).
		\bibitem{asjad} M. Asjad, J. Li, S. Y. Zhu and J. Q. You, Fund. Res. {\bf 3}, 3-7 (2023).
		\bibitem{9} M. Amazioug, S. Singh, B. Teklu and M. Asjad, Entropy {\bf 25}, 1462 (2023).
		\bibitem{SUllah} S. Ullah, H. S. Qureshi, G. Tiaz, F. Ghafoor and F. Saif, Appl. Opt. {\bf 58}, 197-204 (2019).
		\bibitem{mabdi} M. Abdi, et al. Phys. Rev. Lett. {\bf 116.23},233604 (2016).
		\bibitem{Jieli} H. Qian et al. Physical Review A, {\bf 109(1)}, 013704 (2024).
		\bibitem{11} G. S. Agarwal and S. Huang, Phys. Rev. A {\bf 81}, 041803 (2010).
		\bibitem{13} Q. Liao, X. Xiao, W. Nie and N. Zhou, Opt. express {\bf 28}, 5288-5305 (2020).
		\bibitem{14} A. H. Safavi-Naeini, T. M. Alegre, J. Chan, M. Eichenfield, M. Winger, Q. Lin and O. Painter, Nature {\bf 472}, 69-73 (2011).
		\bibitem{15} S. Weis, R. Rivière, S. Deléglise, E. Gavartin, O. Arcizet, A. Schliesser and T. J. Kippenberg, Sci. {\bf 330}, 1520-1523 (2010).
		\bibitem{amghar} M. Amghar, N. Chabar and M. Amazioug. arXiv preprint arXiv:2311. 17731 (2023).
		\bibitem{16} J. P. Marangos, J. Mod. Opt. {\bf 45}, 471-503 (1998).
		\bibitem{17} C. Jiang, H. Liu, Y. Cui, X. Li, G. Chen and B. Chen, Opt. express {\bf 21}, 12165-12173 (2013).
		\bibitem{19} Y. Han, J. Cheng and L. Zhou, J. Phys. B: At. Mol. Opt. Phys. {\bf 44}, 165505 (2011).
		\bibitem{20}K. H. Gu, D. Yan, X. Wang, M. L. Zhang and J. Z. Yin, J. Phys. B: At. Mol. Opt. Phys. {\bf 52}, 105502 (2019).
		\bibitem{21} S. C. Wu, L. G. Qin, J. Jing, T. M. Yan, J. Lu and Z. Y. Wang, Phys. Rev. A {\bf 98}, 013807 (2018).
		\bibitem{22} X. Li, W. Nie, A. Chen and Y. Lan, Phys. Rev. A {\bf 98}, 053848 (2018).
		\bibitem{40} U. Fano, Phys. Rev. {\bf 124}, 1866 (1961).
		\bibitem{400} F. Zangeneh-Nejad and R. Fleury, Phys. Rev. Lett. {\bf 122}, 014301 (2019).
		\bibitem{038} M. V. Rybin, A. B. Khanikaev, M. Inoue, K. B. Samusev, M. J. Steel, G. Yushin and M. F. Limonov, Phys. Rev. Lett. {\bf 103}, 023901 (2009).
		\bibitem{41} K. Qu and G. S. Agarwal, Phys. Rev. A {\bf 87}, 063813 (2013).
		\bibitem{308} Y. F. Xiao, M. Li, Y. C. Liu, Y. Li, X. Sun and Q. Gong, Phys. Rev. A {\bf 82}, 065804 (2010).
		\bibitem{36} X. Zhang, C. L. Zou, L. Jiang and H. X. Tang, Adv {\bf 2}, e1501286 (2016).
		\bibitem{39} X. Zhang, C. L. Zou, L. Jiang and H. X. Tang, Sci. Adv. {\bf 2}, e1501286.
		\bibitem{p}  H. Huebl, C. W. Zollitsch, J. Lotze, F. Hocke, M. Greifenstein, A. Marx and S. T. Goennenwein, Phys. Rev. Lett. {\bf 111}, 127003 (2013).
		\bibitem{q} D. Zhang, X. M. Wang, T. F. Li, X. Q. Luo, W. Wu, F. Nori and J. Q. You, npj Quantum Inf. {\bf 1}, 1-6 (2015).
		\bibitem{23}Y. Tabuchi, S. Ishino, A. Noguchi, T. Ishikawa, R. Yamazaki, K. Usami and Y. Nakamura, Science {\bf 349}, 405-408 (2015).
		\bibitem{25} O. O. Soykal and M. E. Flatté, Phys. Rev. Lett. {\bf 104}, 077202 (2010).
		\bibitem{26} C. A. Potts, E. Varga, V. A. Bittencourt, S. V. Kusminskiy and J. P. Davis, Phys. Rev. X {\bf 11}, 031053 (2021).
		\bibitem{27}X. Zhang, N. Zhu, C. L. Zou and H. X. Tang, Phys. Rev. Lett. {\bf 117}, 123605 (2016).
		\bibitem{28} J. Li, S. Y. Zhu and G. S. Agarwal, Phys. Rev. Lett. {\bf 121}, 203601 (2018).
		\bibitem{00} K. Ullah, M. T. Naseem and Ö. E. Müstecaplıoğlu, Phys. Rev. A {\bf 102}, 033721 (2020).
		\bibitem{29} Y. P. Wang, G. Q. Zhang, D. Zhang, T. F. Li, C. M. Hu and J. Q. You, Phys. Rev. Lett {\bf 120}, 057202 (2018).
		\bibitem{30} R. C. Shen, J. Li, Z. Y. Fan, Y. P. Wang and J. Q. You, Phys. Rev. Lett. {\bf 129}, 123601 (2022).
		\bibitem{31} X. Li, W. X. Yang, T. Shui, L. Li, X. Wang and Z. Wu, J. Appl. Phys. {\bf 128}, (2020).
		\bibitem{32} C. Kong, B. Wang, Z. X. Liu, H. Xiong and Y. Wu, Opt. Express {\bf 27}, 5544-5556 (2019).
		\bibitem{33} Z. X. Yang, L. Wang, Y. M. Liu, D. Y. Wang, C. H. Bai, S. Zhang and H. F. Wang, Front. Phys. {\bf 15}, 1-10 (2020).
		\bibitem{34} A. Kani, B. Sarma and J. Twamley, Phys. Rev. Lett. {\bf 128}, 013602 (2022).
		\bibitem{35} X. Zhang, C. L. Zou, L. Jiang and H. X. Tang, Phys. Rev. Lett. {\bf 113}, 156401 (2014).
		\bibitem{37} W. Qiu, X. Cheng, A. Chen, Y. Lan and W. Nie, Phys. Rev. A {\bf 105}, 063718 (2022).
		\bibitem{38} H. Qian, Z. Y. Fan and J. Li, Quantum Sci. Technol. {\bf 8}, 015022 (2022).
		\bibitem{180} D. F. Walls and G. J. Milburn, Quantum Optics (Springer-Verlag, Berlin, 1994).
		\bibitem{42} K.Ullah, H. Jing and F. Saif, Phys. Rev. A {\bf 97}, 033812 (2018).
		\bibitem{43} K. A. Yasir and W. M. Liu, Sci. Rep. {\bf 6}, 22651 (2016).
		\bibitem{44} K. Ullah, Eur. Phys. J. D. {\bf 73}, 1-9 (2019).
		\bibitem{45} S. Zhang, J. Li, R. Yu, W. Wang and Y. Wu, Sci. Rep. {\bf 7}, 39781 (2017).
		\bibitem{You16}
		Y. P. Wang, G. Q. Zhang, D. Zhang, X. Q. Luo, W. Xiong, S. P. Wang and J. Q. You, Phys. Rev. B {\bf 94}, 224410 (2016).
		
		
	\end{thebibliography}
\end{document}